\documentstyle[multicol,aps,epsfig]{revtex}

\begin{document}
\draft
\title{Elasticity in strongly interacting soft solids: polyelectrolyte network}
\author{ J. Wilder\footnote{e-mail: wilder@mpip-mainz.mpg.de} and T.A. Vilgis \footnote{e-mail: vilgis@mpip-mainz.mpg.de}}
\address{Max-Planck-Institut f\"ur Polymerforschung,
Postfach 3148, 55021 Mainz, Germany}
\date{\today}
\maketitle

\begin{abstract}
  This paper discusses the elastic behavior of a very long crosslinked
  polyelectrolyte chain (Debye-H\"uckel chain), which is weakly charged.
  Therefore the response of the crosslinked chain (network) on an external
  constant force $f$ acting on the ends of the chain is considered. A
  selfconsistent variational computation of an effective field theory is
  employed. It is shown, that the modulus of the polyelectrolyte network has
  two parts: the first term represents the usual entropy elasticity of
  connected flexible chains and the second term takes into account the
  electrostatic interaction of the monomers. It is proportional to the squared
  crosslink density and the Debye - screening parameter.
\end{abstract}



\section{Introduction}

Polyelectrolytes are of fundamental importance in a wide range of academic
sciences to  
applications. It ranges from life sciences such as 
biology or biochemistry to
industrial and practical applications in daily life products. A typical
example for the
latter are
superabsorber materials. These  consist
of highly crosslinked
polyelectrolyte networks which are  strongly
interacting elastic materials. 

The theoretical interest in polyelectrolytes reaches back to the early days of
polymer science (see e.g \cite{tanford}). Nevertheless they belong to the
least understood systems in macromolecular science \cite{schmitz}, since there
are difficulties to apply renormalization group theories and scaling ideas in
which long ranged (i.e. Coulomb) forces are present. Only very recently novel
types of field
theoretic attempts brought progress \cite{liverpool}.

In the present paper we aim for a theory of the elasticity of polyelectrolyte
networks. This is a non trivial task, since most of the classical and modern
theories neglect the effect of interactions on elasticity.  In neutral
networks the interactions are mainly given by excluded volume forces. In the
dry network state these can be safely neglected for most cases, since in such
dense systems like polymer melts excluded volume interactions are screened
largely \cite{edwards:75,doied}. However, if solvent is added to the network
and the network starts to swell, problems arises. Early theories by Flory
\cite{flory} suggested that the elastic part of the free energy and the
solvent part, i.e., a Flory Huggins type term, can be added. Later on this
concept has been named the Flory - Rehner - Hypothesis \cite{fh} in the
context of swelling experiments. The network state would be determined by the
minimum of the total free energy. Indeed such approximations are used in a
wide range of application for rubberlike materials in the swollen state
\cite{eichinger}. The comparison with experiments seem to be reasonable, i.e.,
in all cases the modulus is found to be proportional to the crosslink density,
although from a theoretical point of view the simple addition of the two parts
of the free energy must be wrong \cite{ball}. It must be wrong, because a
complete formulation of the partition function suggests immediately the
appearance of cross terms. In a subsequent paper \cite{preprint}, we will
analyse the Flory - Rehner - Hypothesis in the context of thermodynamics and
its application. There it is shown that corresponds to an approximation which
neglect fluctuations completely. In the present paper this is not the central
point and we restrict ourselves to compute the elastic response of a
polyelectrolyte network including fluctuations on a variational level.

In neutral networks, however the approximations
seem to be not too bad, because the interactions are relatively short ranged
and weak. Moreover the equilibrium swelling degree is then given 
by the $c^{*}$ - network. When  polyelectrolyte networks are considered, we
cannot expect that the Flory approximation holds. The interactions are long
ranged and very strong compared to excluded volume interactions. Here a
strong interplay of elastic degrees of freedom and interactions must be
expected. The reason is very simple: The strong interactions change the
physical nature and the conformation of a charged polymer chain strongly
compared to the neutral chain. The state at rest, i.e., a network strand
without application of an external force does not contain as many degrees of
conformation as the equivalent neutral chain. Its conformation ranges, 
depending on ionization
and salt content of the solution, from excluded volume behavior to a rod like
behavior. Thus a more detailed theory is needed to compute the elastic
modulus of charged and highly interacting gels. 
Nevertheless the Flory  approximation has been
employed also for strongly interacting polyelectrolyte gels in bad solvent to
study the phase diagram \cite{khokhlov}. Moreover also recent suggestions
\cite{muthy} have claimed an unchanged modulus for polyelectrolyte
networks. We will see later, however, the reasons for these statements.

Most of the classical network theories rely on "single chain models". This is
to say that the elasticity of the network can be roughly computed by studying
first the elasticity of a single chain. The elastic properties of the entire
network is then supposed to be given by the partition function of the single
chain raised to the power of the number of chains. Such ways of computations
hold strongly only for weakly interacting systems. Again in neutral networks
the interactions, i.e., the excluded volume forces, are weak since they are
screened, but in polyelectrolyte networks these type of assumptions do not
seem to be reasonable. Another drawback of these single chain type theories
is that they ignore the effect of quenched disorder stemming from the
crosslinks completely. Although some progress for polyelectrolyte networks has
been achieved by single chain theories \cite{rubi}, which are based on blob
pictures \cite{degennesblob}, it is necessary to study a network model
including the quenched disorder produced by the crosslinks (see e.g.
\cite{edvil} for a review). It has to be kept in mind that so so far only
the limits for large and small screenig parameters are studied. Large screenig
parameters (in terms of a Debye - H\"uckel approximation)
correspond to a large salt concentration. This regime is relatively
unintersting, since it corresponds to the good solvent regime. For small
screening the chains are significantly stretched, and the chains are in the
polyelectrolyte regime \cite{degennesblob}. For networks, however, these two
limits allone are not sufficient. In strongly crosslinked networks the chain
pieces between two crosslinks can easily be of the same order as the 
Debye - H\"uckel screening length. Therefore a more elaborated analysis must
be carried out.

In a previous  
paper \cite{HaWiVi} we had investigated the coupling between
elasticity and interactions in a single chain. This model calculation was
carried out to
show that a strong coupling between elastic (conformational) degrees of freedom
and electrostatic interactions exists. Of course, unlike as in neutral network
systems it is not sufficient to consider  the elasticity of a single chain
and generalize the results to a corresponding network of a large number of
such chains. In the case of polyelectrolyte networks the chains are strongly
interacting with each other. Therefore any assumption on weakly interacting
chains fails. Nevertheless the computational method used in \cite{HaWiVi} has
shown to be useful. The results presented there ended up in a two regime 
blob picture. For small forces the de Gennes \cite{degennesblob}
electrostatic blob model was
recovered. In this regime the chain was relatively easy to deform.
For larger forces a change in the elasticity was predicted. Then
the chain entered in a Pincus regime of the prestretched chain. The results
have been confirmed by simulation and by a self consistent variational
principle. Indeed the method developed in \cite{HaWiVi} is appropraite to the
intermediate regime between strong and weak screening.

In this paper we extend our considerations concerning the single
polyelectrolyte chain \cite{HaWiVi} to a polyelectrolyte network. 
The simplest version of a polymer network was introduced by Deam and Edwards
\cite{rep}, which
consists of a very (macroscopically) long  crosslinked chain. 
In our case there is no difference
between a network made of many chains or one long chain, since we assume to be
deep in the solid phase, i.e., well beyond the vulcanization threshold. 
As in the case of a single polyelectrolyte chain \cite{HaWiVi} we are
interested in the force-size relationship of the network in a good solvent.  
In contrast to the classical  theories the effects of 
interactions are now taken into account explicitly. For simplicity we assume
 a Debye-H\"uckel potential for the electrostatic interaction. It is of the
form $V({\bf r}) \propto 1/r \exp(-r/\lambda)$.
This might not be the best choice to reproduce recent
simulation data \cite{Krems} which had shown that the Debye-H\"uckel potential
is not always a good approximation, but the advantage of the potential is that
its range is controlled by a single parameter, i.e., 
the Debye screening length $\lambda$. Indeed little is known about the 
interplay of elasticity (conformation) and strength of the interactions
in the theory of network elasticity. The aim of the present  paper is to learn
something about this interplay. Therefore we apply an external force on the
ends of the chain. To do so, we employ a variational principle 
and determine the effective
propagator of the chain, which allows statements about the interplay  of 
conformation
{\bf and} interactions on the elasticity. In the following we will neglect the
effect of entanglements completely, since we are mainly interested in the
contributions of the interactions to elasticity. 
The results given below will then mainly apply for unentangled networks.
 To produce simple estimates
of entanglement contributions tube like models can be employed
\cite{edvil}. We expect, however, that the effects of entanglements will not be
very different from those in neutral networks, as long as their number is
given. The main problem would then be to compute the mean number of
entanglements by the presence of the electrostatic interactions.

The starting point for the computation of the elastic 
free energy is the Green function of the crosslinked chain without an
external force, similar to our calculations of the single polyelectrolyte
chain \cite{HaWiVi}. The force is treated through the analytic continuation of
the Fourier transformed Greens function to the complex plane. After having
introduced a field theory the problem is mapped on
a Gaussian field theory with a propagator that formally in the Fourier space
can be written down exactly by making use of the proper self energy. According
to the well known Feynman variational inequality the 
sum of the Gaussian free energy and 
the mean-value of the interacting potential has to be minimized with respect 
to the proper self energy, which is our variational parameter. This leads to 
a non-linear integral equation for the proper self energy, which can be solved
 approximately.

The paper is organized as follows. In the next section we present the
underlying model and introduce the crosslinks. In section III we formulate a
field theory and calculate the variational equation for the proper self
energy. In section IV this variational equation is solved approximately. These
sections will be written out in more detail, since this is - to our knowledge
- the first time that a strongly interacting network is investigated by this
technique. Thus the mathematically interested reader might find the main steps
of the computation. The
results are presented in section V. The paper ends with the discussion 
of the results.

\section{Model and theoretical background}
The starting point for the field theoretic computation is a network 
formed out of a macroscopically long chain by the instantaneous introduction
of a sufficiently large number of
crosslinks in the liquid phase (see for example \cite{rep,PanRab}).
We restrict ourselves to a network, which consists of flexible, weakly charged 
strands. Consequently the Edwards model is an appropriate tool to describe the
network \cite{Ledwards}. Therefore let us choose as Hamiltonian for 
 the charged chain in aqueous
solution:
\begin{eqnarray}
&\displaystyle\label{model1}   \beta H_{\rm E} [{\bf r};{\bf f}] = {3 \over 2l^{2} } \int^{N_{\rm tot}}_{0} \mbox{d} s \
\left({\mbox{d} {\bf r} \over \mbox{d} s}\right)^2 + \beta
 \int^{N_{\rm tot}}_{0} \mbox{d} s \  {\bf f} \cdot
 {\mbox{d} {\bf r} \over \mbox{d} s}
& \nonumber \\&\displaystyle
 + \
\frac{bz^{2}}{2} \int^{N_{\rm tot}}_{0} \mbox{d} s \int^{N_{\rm tot}}_{0} \mbox{d} s' \
{\exp\left\{-\kappa\left \vert {\bf r} (s) - {\bf r} (s') \right \vert \right \}
\over \left \vert {\bf r} (s) - {\bf r} (s') \right \vert}\,,
\end{eqnarray}
where ${\bf r}(s)$ represents the 
chain conformation in three dimensions as a function of the
contour variable $s$, $b=e^{2}/4\pi
\epsilon_{0}\epsilon_{\rm r}k_{\rm B}T$ is the Bjerrum-length, $\beta$ is
$(k_{\rm B}T)^{-1}$, where $k_{\rm B}$ is the Boltzmann constant and $T$ 
denotes the absolute temperature. $l$ is the Kuhn segment length,
$z$ is the monomer charge in units of $e$, $\epsilon_{0}$ is the dielectric
constant and $\epsilon_{\rm r}$ the relative dielectric constant. $N_{\rm tot}$ 
stands for the bare number of
monomers on the chain, ${\bf f}$ is the external force and $\kappa^{-1}$
denotes the Debye-H\"uckel screening length $\lambda$. For the introduction of
crosslinks
we choose the standard way suggested by Deam and Edwards \cite{rep}. We
assume for mathematical convenience four functional crosslinks which join
two arbitrary segments ${\bf r}(s_{i})$ and ${\bf r}(s_{j})$ along the
chain.
Of course, the value for the free energy then depends on the specific choice of
the pairs of monomers, but on macroscopic scale only the statistical average
on any crosslink configuration
is of importance. Nevertheless this requires non - Gibbsian statistical
mechanics in the sense that the crosslink positions represent quenched degrees
of freedom.

The basic problem for the determination of the free energy of the network is
the presence of quenched disorder, which is contained in the permanent
crosslinks. The formation of a crosslink, i.e., the linkage between two
arbitrary segments ${\bf r}(s_{i})$ and ${\bf r}(s_{j})$ represents quenched
disorder, since the segments are joined together for all times and for all
thermodynamic situations. Of course, the experimental relevant free energy $F$
depends on the crosslink configuration and crosslink realization ${\bf S}$.
The actual crosslink configuration ${\bf S}$ is not known in detail, thus the
technical difficulty is to average the free energy $F(\bf S)$ over all
possible crosslink realizations. To do so, it is generally assumed that the
corresponding distribution ${\cal P}({\bf S})$ can be determined.

Let us perform the outline of the idea in more detail. To
calculate the free energy $F$ of the network, we have to take the statistical
average over all crosslink configurations ${\bf S}$. This represents the fact,
that $F$ is a self averaging quantity.
\begin{eqnarray}
F(N_{\rm tot},N_{\rm c})=-k_{\rm B}T\int {\mbox d}{\bf S}\, {\cal P}({\bf S})\ln Z({\bf
  S}).\label{m2}
\end{eqnarray}
$Z({\bf S})$ is the constrained partition function for a network with
the crosslink configuration ${\bf S}$, $N_{\rm c}$ is the number 
of crosslinks and ${\cal P}({\bf S})$ is the crosslink
distribution function. Since we assume that the crosslinks are instantaneously
introduced in the liquid phase, ${\cal P}({\bf S})$ is yielded by the
constrained partition function of the liquid phase, which is defined in terms
of a path integral as:
\begin{eqnarray}
Z^{(0)}({\bf S})=\int {\cal D}{\bf r}(s)\exp(-\beta
H_{\rm E})\prod_{(i,j)}\delta[{\bf r}(s_{i})-{\bf r}(s_{j})],\label{m3}  
\end{eqnarray}
where $(i,j)$ denotes that the $i$-th and the $j$-th monomer are close to each
other in the liquid phase, which means that they can form 
one of the $N_{\rm c}$
crosslinks. 
Consequently the crosslink distribution function ${\cal P}({\bf S})$
is simply:
\begin{eqnarray}
{\cal P}({\bf S})=\frac{Z^{(0)}({\bf S})}{\int {\mbox d}{\bf S}^{'} Z^{(0)}({\bf
  S}^{'})}\,.\label{m4}
\end{eqnarray}
In the following it appears to be reasonable to 
assume that the so chosen distribution function does not depend on the specific
deformation of the network. 
Note that $Z(\bf S)$ differs generally
from $Z^{(0)}({\bf S})$. Since we are interested
in deformations of the network,  $Z(\bf S)$ is the partition function of the
deformed network.

To calculate the free energy $F$ (Eq. (\ref{m2})) explicitly it is convenient
to make use of the so called replica
trick \cite{rep}. This, so far, purely mathematical  trick relies on the
identity 
$$
\ln z = \frac{\partial z^{m}}{\partial m}{\Bigg\vert}_{m=0}
$$ 
Define
\begin{eqnarray}
F_{m}(N_{\rm tot},N_{\rm c})=-k_{\rm B}T\ln 
\int {\mbox d}{\bf S}\,Z^{(0)}({\bf
S})Z^{m}({\bf S})\,,\label{m5}
\end{eqnarray}
where $m$ is the replica index. Eq. (\ref{m5}) shows the origin of the
technical term "replica method". By the use of the mathematical trick $m$
copies of the system are produced.
The free energy $F$, which is averaged  
over the disorder of the crosslinks, reads \cite{rep}:
\begin{eqnarray}
F(N_{\rm tot},N_{\rm c})=\frac{{\partial}F_{m}(N_{\rm tot},N_{\rm
    c})}{{\partial }m} {\Bigg\vert}_{m=0} \label{ef6}
\end{eqnarray}
As in the previous  paper, the free energy $F$ is calculated by making use of
its relation to the corresponding distribution functions and Green functions
of the corresponding propagator (see \cite{HaWiVi} for the technical details).

\section{Field-theoretical formulation}

In the following we will give an outline of the computation of the network
elasticity, i.e., our main aim is to compute the low deformation modulus of the
polyelectrolyte network. Therefore we start from a concentrated
polyelectrolyte solution, consisting of one macroscopic chain and the
appropriate number of counterions to satisfy the condition of
electro-neutrality. Then the crosslinks are introduced instantaneously by the
process described above. The following chapter will be very formal, but we
think that it is important to do so, since it turned out that 
non of the methods employed for neutral networks can be used in the present 
context. The main reason for this is that here we do not have the option of
formulating the problem in terms of one length scale, i.e., the meshsize, but
must take into account the range of the interaction.  

For the relevant quantities to compute it is necessary to
consider the  correlation function $G({\bf \hat{r}},N_{\rm
  tot},N_{\rm c},{\bf f},{\bf S})$ of a crosslinked polyelectrolyte chain in
replica space, where ${\bf
  \hat{r}}=({\bf r}_{0},{\bf r}_{1},\dots,{\bf r}_{m})$ is the replicated
$3(m+1)$-dimensional end-to-end vector of the chain and ${\bf S}$ the specific
crosslink configuration:
\begin{eqnarray}
G({\bf \hat{r}},N_{\rm tot},N_{\rm c},{\bf f},{\bf S})&=&\int_{{\bf r}_{0}(0)=0}^{{\bf
    r}_{0}(N_{\rm tot})={\bf r}_{0}}{\cal D}{\bf r}_{0}(s)\int_{{\bf r}_{1}(0)=0}^{{\bf
    r}_{1}(N_{\rm tot})={\bf r}_{1}}{\cal D}{\bf r}_{1}(s)\dots\int_{{\bf r}_{m}(0)=0}^{{\bf
    r}_{m}(N_{\rm tot})={\bf r}_{m}}{\cal D}{\bf r}_{m}(s)\nonumber \\
&\times& \exp(-\beta {\hat H}_{\rm E}[{\bf {\hat r}},{\bf
  f}])\prod_{p=0}^{m}\prod_{(i,j)}\delta [{\bf r}_{p}(s_{i})-{\bf r}_{p}(s_{j})]\label{ft0}
\end{eqnarray}
with $p$ the replica index. The force ${\bf f}$ is only acting on the ends of
the final states (replica index 1 to $m$) of the chain. Thus the replicated
Hamiltonian ${\hat H}_{\rm E}[{\bf {\hat r}},{\bf f}]$ reads
\begin{eqnarray}
\beta {\hat H}_{\rm E}[{\bf {\hat r}},{\bf f}])&=& \frac{3}{2l^{2}}\sum_{p=0}^{m}
 \int^{N_{\rm tot}}_{0} \mbox{d} s \
\left(\frac{\mbox{d} {\bf r}_{p}}{\mbox{d} s}\right)^2 + \beta \sum_{p=1}^{m} 
\int_{0}^{N_{\rm tot}}\mbox{d} s\, {\bf f}\frac{\mbox{d}{\bf
    r}_{p}}{\mbox{d}s}\nonumber \\
&+&\frac{bz^{2}}{2} \sum_{p=0}^{m}  \int^{N_{\rm tot}}_{0} \mbox{d} s \int^{N_{\rm tot}}_{0}
\mbox{d} s' \frac{\exp\left\{-\kappa\left \vert {\bf r}_{p} (s) - {\bf r}_{p}
      (s') \right \vert \right\}}{ \left \vert {\bf r}_{p} (s) - {\bf r}_{p}
    (s') \right \vert }\label{ft01} 
\end{eqnarray}
The important observation is that the replicated Hamiltonian $\hat H_{\rm E}$
separates in the different replicas. The coupling of the replicas comes into
play when the average over the distribution ${\cal P}({\bf S})$ is
performed. If we use the standard distribution \cite{rep} the Green function
must be computed upon the effective Hamiltonian
\begin{equation}
{\hat H} = {\hat H}_{\rm E} - z_{\rm c}\int_{0}^{N_{\rm tot}} {\text d}s
\int_{0}^{N_{\rm tot}} {\text d}s'\prod_{p=0}^{m}
\delta({\bf r}_{p}(s) - {\bf r}_{p}(s'))
\end{equation}
 The latter equation shows
the difficulty of the problem, i.e., all replicas are coupled. Below we choose a
different way as suggested by Edwards \cite{rep} and Panyukov \cite{PanRab}.
We must employ field theoretic methods ( as also done in \cite{PanRab}), but
the treatment of the field theory is very different, since the symmetry of
the problem is not of the same nature as in the case of neutral networks. 

It is easy to show that the Greens function in the Fourier transformed
replica-space depending on a constant force ${\bf f}$ can be calculated by a zero
force Greens function. The force can be reintroduced by the analytic
continuation of the Fourier-space to the complex plane, which means in detail:
\begin{eqnarray}
G({\bf \hat{k}},N_{\rm tot},N_{\rm c},{\bf
  f},{\bf S})&=&\int \mbox{d}{\bf {\hat r}}\,\exp\{-i({\bf {\hat k}}-i\beta {\bf
  {\hat f}}){\bf {\hat r}}\} G({\bf \hat{r}},N_{\rm
  tot},N_{\rm c},{\bf f}={\bf 0},{\bf S})\nonumber \\
&=&G({\bf k}^{(0)},{\bf k}^{(1)}-i\beta {\bf f},\dots,{\bf k}^{(m)}-i\beta
{\bf f},N_{\rm tot},N_{\rm c},{\bf f}={\bf 0},{\bf S})\label{ft1a}
\end{eqnarray}
Here ${\bf {\hat f}}$ is the $3(m+1)$-dimensional force vector $({\bf 0},{\bf
  f},\dots,{\bf f})$. 
This is exactly the same mathematical  procedure
as we had already used in the previous  calculation concerning
the single polyelectrolyte chain (see \cite{HaWiVi}). As a consequence of
Eq. (\ref{ft1a}) in the following we neglect the force term in the Hamiltonian
and first calculate a zero force correlation function.

The grand canonical correlation function $\tilde{G}({\bf
   \hat{k}},\mu_{0},z_{\rm c},{\bf f})$ in replicated Fourier space, 
where ${\bf
  \hat{k}}$ is the wave vector in the $3(m+1)$-dimensional Fourier transformed
  replica space, $\mu_{0}$
is the chemical potential of the monomers and $z_{\rm c}$ 
is the fugacity of the
crosslinks, can be calculated by the introduction of de Gennes'
zero-component field theory (see for example \cite{PanRab})
\begin{eqnarray}
\tilde{G}({\bf {\hat k}},\mu_{0},z_{\rm c},{\bf f})=\lim_{n \to 0}\int{\cal D}\vec{\psi}\,\psi_{1}({\bf
  {\hat k}})\psi_{1}(-{\bf {\hat k}})\exp\{-\beta H[\vec{\psi}]\}.\label{ft1}
\end{eqnarray}
$H[\vec{\psi}]$ in Eq. (\ref{ft1}) is the zero force field theoretical Hamiltonian expressed by the
$n$-component field $\vec{\psi}$, which in Fourier space
reads \cite{HaWiVi,PanRab}:
\begin{eqnarray}
H[\vec{\psi}({\bf \hat{q}})]&=&\int_{{\bf
    \hat{q}}}\left[\frac{\mu}{2}\vec{\psi}({\bf \hat{q}})\vec{\psi}(-{\bf
    \hat{q}})+\frac{l^{2}}{2}{\bf \hat{q}}^{2}\vec{\psi}({\bf
    \hat{q}})\vec{\psi}(-{\bf \hat{q}})\right]\nonumber \\
&-&\frac{z_{\rm c}}{8}\int_{{\bf \hat{q}}_{1},{\bf \hat{q}}_{2},{\bf
    \hat{q}}_{3},{\bf \hat{q}}_{4}} \vec{\psi}({\bf
  \hat{q}}_{1})\vec{\psi}({\bf \hat{q}}_{2})\vec{\psi}({\bf
  \hat{q}}_{3})\vec{\psi}({\bf \hat{q}}_{4})\delta({\bf \hat{q}}_{1}+{\bf
  \hat{q}}_{2}+{\bf \hat{q}}_{3}+{\bf \hat{q}}_{4})\nonumber \\
&+&\sum_{k=0}^{m}\left[\int_{{\bf \hat{q}}_{1},{\bf
      \hat{q}}_{2}}\vec{\psi}({\bf \hat{q}}_{1})\vec{\psi}({\bf \hat{q}}_{2})\prod_{l\neq
    k}\delta({\bf q}_{1}^{(l)}+{\bf q}_{2}^{(l)})\right]\label{ft2} \\
&\times&\left[\int_{{\bf \hat{q}}_{3},{\bf \hat{q}}_{4}}\vec{\psi}({\bf
    \hat{q}}_{3})\vec{\psi}({\bf \hat{q}}_{4})\prod_{l\neq
    k}\delta({\bf q}_{3}^{(l)}+{\bf q}_{4}^{(l)})\right]V^{(k)}({\bf
  q}_{3}^{(k)}+{\bf q}_{4}^{(k)})\delta({\bf q}_{1}^{(k)}+{\bf
  q}_{2}^{(k)}+{\bf q}_{3}^{(k)}+{\bf q}_{4}^{(k)})\nonumber
\end{eqnarray}
where $V^{(k)}(q)$ is the Fourier transform of the Debye-H\"uckel potential in
the $k$-th replica, $\vec{\psi}$ is a $n$-component vector field and 
$\int_{\bf q}$ is an abbreviation for the integral notation 
$\int {\mbox d}^{d}q/(2\pi)^{d}$
with $d$ the dimension of the vector ${\bf q}$. 
In Fourier space $\tilde{G}({\bf \hat{k}},\mu_{0},z_{\rm c})$ can be written
exactly as:
\begin{eqnarray}
\tilde{G}({\bf \hat{k}},\mu_{0},z_{\rm c})=\left(\mu_{0} +\frac{l^{2}}{6}{\bf
    \hat{k}}^{2}+\Sigma({\bf \hat{k}},z_{\rm c})\right)^{-1}\label{ft3}
\end{eqnarray}
where $\Sigma({\bf \hat{k}},z_{\rm c})$ denotes the proper self energy in replica
space.

Since we do not know the exact proper self energy $\Sigma({\bf
  \hat{k}},z_{\rm c})$, we have to calculate it approximately. Therefore
we now consider an approximate correlation function $\tilde{{\cal G}}({\bf
  \hat{k}},\mu_{0},z_{\rm c})$ with an approximate proper self-energy $ M({\bf
  \hat{k}},z_{\rm c})$: 
\begin{eqnarray}
\tilde{{\cal G}}({\bf \hat{k}},\mu_{0},z_{\rm c})=\left(\mu_{0}
  +\frac{l^{2}}{6}{\bf \hat{k}}^{2}+M({\bf \hat{k}},z_{\rm c})\right)^{-1}\label{ft4}
\end{eqnarray}
To proceed with a variational principle we define the Gaussian Hamiltonian ${\cal H}$ by
\begin{eqnarray}
\beta {\cal H}[\vec{\psi}]=\frac{1}{2}\int_{{\bf \hat{k}}}\vec{\psi}(-{\bf \hat{k}})\tilde{{\cal
    G}}^{-1}({\bf
  \hat{k}},\mu_{0},z_{\rm c})\vec{\psi}({\bf \hat{k}})\label{ft5}
\end{eqnarray}
The correlation function $\tilde{{\cal G}}({\bf \hat{k}},\mu_{0},z_{\rm c})$ can
be calculated within the zero-component field theory of de Gennes (see for
example \cite{PanRab}):
\begin{eqnarray}
\tilde{{\cal G}}({\bf \hat{k}},\mu_{0},z_{\rm c})=\lim_{n \to 0}\int{\cal D}\vec{\psi}\,\psi_{1}({\bf
  \hat{k}})\psi_{1}(-{\bf \hat{k}})\exp\{-\beta {\cal H}[\vec{\psi}]\}\label{ft6}
\end{eqnarray}
In this notation the well-known Feynman inequality, which can be taken for the
calculation of the approximate proper self energy $M({\bf \hat{k}},z_{\rm c})$, is given by:
\begin{eqnarray}
F\le {\cal F}+\langle H-{\cal H}\rangle_{{\cal H}}\label{ft7}
\end{eqnarray}
where
\begin{eqnarray}
\langle\dots\rangle_{{\cal H}}=\lim_{n \to 0}\frac{\int{\cal
    D}\vec{\psi}\,\dots\exp\{-\beta {\cal H}\}}{\int{\cal
    D}\vec{\psi}\,\exp\{-\beta {\cal H}\}}\label{ft8}
\end{eqnarray}
is the mean-value and ${\cal F}$ the free energy with respect to ${\cal H}$.
 The right hand side of the inequality
(\ref{ft8}) has to be minimized with respect to 
$M$. ${\cal F}$ and $\langle H-{\cal H}\rangle_{{\cal H}}$ can be written 
in terms of the correlation function $\tilde{{\cal G}}({\bf
  \hat{k}},\mu_{0},z_{\rm c})$: 
\begin{equation}
\beta{\cal F}=-\frac{n}{2}V\int_{{\bf \hat{q}}}\,
\ln\left[\tilde{\cal G}({\bf \hat{q}},\mu,z_{\rm c})\right]\label{ft10}
\end{equation}
where $V$ is the volume of the replica space.
As can be shown easily the second term of the right hand side of inequality
(\ref{ft7}) is \cite{Ma}
\begin{eqnarray}
\beta<H-{\cal H}>_{\cal H}&=&-\frac{n}{2}V\int_{{\bf \hat{q}}} \,M({\bf
  \hat{q}},z_{\rm c})\tilde{\cal G}({\bf \hat{q}},\mu,z_{\rm c}))+\frac{\pi
  bz^{2}n^{2}(m+1)V_{0}^{m}V}{2\kappa^{2}}\left(\int_{{\bf
      \hat{q}}}\,\tilde{\cal G}({\bf \hat{q}},\mu,z_{\rm c})\right)^{2}\nonumber \\
&+&\pi bz^{2}n(m+1)V\int_{{\bf \hat{q}}_{1},{\bf \hat{q}}_{2}}
\,\frac{\tilde{\cal G}({\bf \hat{q}}_{1},\mu,z_{\rm c})\tilde{\cal G}({\bf
    \hat{q}}_{2},\mu,z_{\rm c})}{({\bf q}_{1}^{(0)}+{\bf
    q}_{2}^{(0)})^{2}+\kappa^{2}}\prod_{l=1}^{m}\delta({\bf q}_{1}^{(l)}+{\bf q}_{2}^{(l)})\nonumber \\
&-&\frac{z_{\rm c}}{8}(n^{2}+2n)V\left(\int_{{\bf \hat{q}}}\,\tilde{\cal
    G}({\bf \hat{q}},\mu,z_{\rm c})\right)^{2}\label{ft9}
\end{eqnarray} 
with $V_{0}$ the volume of a single replica segment and $V$ the volume of the
whole replica-space. As we want to determine the approximate proper self
energy $M({\bf \hat{k}},z_{\rm c})$, this function should be the
variational parameter. Consequently the general minimization condition reads:
\begin{eqnarray}
\frac{\delta}{\delta  M({\bf
    \hat{q}},z_{\rm c})}({\cal F}+\langle H-{\cal H}\rangle_{{\cal H}})=0\label{ft10}
\end{eqnarray}
where $\delta/\delta M({\bf \hat{q}},z_{\rm c})$ denotes the functional derivative with
respect to $ M({\bf \hat{q}},z_{\rm c})$. After inserting Eqs. (\ref{ft8}) and
(\ref{ft9}) into Eq. (\ref{ft10}) one obtains 
\begin{eqnarray}
M({\bf \hat{k}},z_{\rm c})&=&\frac{2\pi
  bz^{2}n(m+1)V_{0}^{m}}{\kappa^{2}}\int_{{\bf
    \hat{q}}}\,\frac{1}{\mu+\frac{l^{2}}{6}{\bf \hat{q}}^{2}+M({\bf
    \hat{q}},z_{\rm c})}\nonumber \\
&+&4\pi bz^{2}\int_{{\bf q}^{(0)}}\,\frac{1}{({\bf q}^{(0)}+{\bf
    k}^{(0)})^{2}+\kappa^{2}}\frac{1}{\mu+\frac{l^{2}}{6}({\bf q}^{(0)},{\bf
    k}^{(1\dots m)})^{2}+M({\bf q}^{(0)},{\bf k}^{(1\dots m)})}\label{ft12}\\
&+&4\pi bz^{2}m\int_{{\bf q}^{(1)}}\,\frac{1}{({\bf q}^{(1)}+{\bf
    k}^{(1)})^{2}+\kappa^{2}}\frac{1}{\mu+\frac{l^{2}}{6}({\bf k}^{(0)},{\bf
    q}^{(1)},{\bf k}^{(2\dots m)})^{2}+M({\bf k}^{(0)},{\bf q}^{(1)},{\bf k}^{(2\dots m)})}\nonumber \\
&+&
\frac{z_{\rm c}}{2}(n+2)\int_{{\bf \hat{q}}}\,\frac{1}{\mu+\frac{l^{2}}{6}{\bf
    \hat{q}}^{2}+M({\bf \hat{q}},z_{\rm c})},\nonumber
\end{eqnarray}
where the short hand notation ${\bf k}^{(i\dots m)}= ({\bf k}^{(i)},\dots
,{\bf k}^{(m)})$ is introduced. This is a non-linear integral equation for $
M({\bf \hat{q}},z_{\rm c})$, which in the following has to be solved
approximately, since the exact solution is unknown.

\section{Approximate solution for the proper self-energy}
In analogy to the calculations on the single chain \cite{HaWiVi} we restrict
ourselves to small external forces applied on the ends of the crosslinked
chain. Therefore we make the same {\it ansatz} for the proper-self energy as in the
case of the single chain \cite{HaWiVi}. The only difference is that the proper
self energy in this paper is a function depending on the replica-space wave
vector ${\bf \hat{q}}$:
\begin{equation}
M({\bf \hat{q}})=a_{0}+a_{1}{\bf \hat{q}}^{2}+{\cal O}({\bf \hat{q}}^{4})\label{l1}
\end{equation}
To start with let  $ M_{\rm r}({\bf \hat{q}})$ be $ M({\bf
  \hat{q}})- M({\bf 0})$. Then $ M_{\rm r}({\bf \hat{q}})$ is given by
\begin{eqnarray}
M_{\rm r}({\bf \hat{k}})&=&4\pi bz^{2}(m+1)\left[\int_{{\bf
      q}^{(0)}}\,\frac{1}{({\bf q}^{(0)}+{\bf
      k}^{(0)})^{2}+\kappa^{2}}\,\frac{1}{\mu_{\rm r}+\frac{l^{2}}{6}({\bf
      q}^{(0)},{\bf k}^{(1 \dots m)})^{2}+M_{\rm r}({\bf q}^{(0)},{\bf k}^{(1 \dots m)}}\right.\nonumber \\
&-&\left. \int_{{\bf q}^{(0)}}\,\frac{1}{({\bf
      q}^{(0)})^{2}+\kappa^{2}}\,\frac{1}{\mu_{\rm r}+\frac{l^{2}}{6}({\bf
      q}^{(0)})^{2}+M_{\rm r}({\bf q}^{(0)},{\bf 0})}\right]\label{l2}
\end{eqnarray}
where $\mu_{\rm r}=\mu+a_{0}$. Since we assume a replica symmetric solution for
the proper-self energy, we consider the second derivative of $M_{\rm r}(\hat{k})$ with
respect to $k^{(0)}$ at vanishing replica-space wave vectors ${\bf
  \hat{k}}$, which yields a result for $a_{1}$. The details of calculation are
exactly the same as for the single chain \cite{HaWiVi}:
\begin{equation}
a_{1}=\frac{2bz^{2}}{3l^{2}\pi \mu_{\rm r} \kappa}
\left [1+{\cal O}\left (\kappa l \over \sqrt{\mu_{\rm r}}  \right)\right]\label{l3}
\end{equation}
This result is valid for $\beta\hat{q}/\kappa<1$ or in terms of the force
$\beta f/\kappa<1$. For details see \cite{HaWiVi}.
It is important to mention that the coefficient $a_{1}$ is independent of the
fugacity of the crosslinks $z_{\rm c}$. Consequently to study the characteristics
of the network it is necessary to calculate the constant term of the
proper-self energy $a_{0}$. In the following we neglect terms of order $n$,
where $n$ is the number of components of the field $\vec{\psi}$, since we have
to take the limit $n\to 0$:
\begin{eqnarray}
M({\bf \hat{k}})&=&4\pi bz^{2}(m+1)\int_{{\bf q}^{(0)}}\frac{1}{({\bf
    q}^{(0)}+{\bf k}^{(0)})^{2}+\kappa^{2}}\,\frac{1}{\mu+\frac{l^{2}}{6}({\bf
    q}^{(0)},{\bf k}^{(1 \dots m)})^{2}+M({\bf q}^{(0)},{\bf k}^{(1 \dots m)})}\nonumber \\
&+&z_{\rm c}\int_{{\bf \hat{q}}}\,\frac{1}{\mu+\frac{l^{2}}{6}{\bf
    \hat{q}}^{2}+M({\bf \hat{q}})}\label{l4}
\end{eqnarray}

Note, that the right hand side of equation (\ref{l4}) diverges even in the
limit $m\to 0$. Since we started with a discrete model, we are not allowed to
consider
infinitesimal small length scales, which means, that we have to introduce a
cutoff in the integrals of equation (\ref{l4}). An appropriate cutoff is
$\kappa$ (see \cite{HaWiVi}). Consequently it is consistent to substitute the
{\it ansatz} (\ref{l1}) for $M({\bf \hat{q}})$ in the integrals of Eq. (\ref{l4}), since
it is valid for $|{\bf \hat{q}}|<\kappa$ \cite{HaWiVi}. As we intend to calculate
$a_{0}$, all $k$ variables in the integrals of Eq. (\ref{l4}) vanish, which
means that only absolute values of the ${\bf q}$ variables occur. Therefore we
introduce spherical coordinates and neglect terms of order ${\cal O}(m^{2})$
in the second integral after having transformed to a dimensionless integration
variable.
Thus we get the following selfconsistent equation for $a_{0}$:
\begin{eqnarray}
a_{0}&=&\frac{2bz^{2}(m+1)}{\pi}\int_{0}^{\kappa} {\mbox
  d}q\frac{q^{2}}{q^{2}+\kappa^{2}}\,\frac{1}{\mu+a_{0}+\frac{l^{2}}{6}q^{2}+a_{1}q^{2}}\nonumber \\
&+&\frac{2z_{\rm c}(\mu+a_{0})^{\frac{3}{2}m+\frac{1}{2}}}{2^{3(m+1)}\pi^{\frac{3}{2}(m+1)}\Gamma\left(\frac{3}{2}(m+1)\right)\left(\frac{l^{2}}{6}+a_{1}\right)^{\frac{3}{2}(m+1)}}\int_{0}^{\kappa\sqrt{\frac{l^{2}/6+a_{1}}{\mu+a_{0}}}}{\mbox
  d}\tilde{q}\frac{\tilde{q}^{2}(1+3m\ln(\tilde{q}))}{1+\tilde{q}^{2}}\label{l5}
\end{eqnarray} 
Note, that the integration-variable $\tilde{q}$ is dimensionless. The
integrations in Eq. (\ref{l5}) can be performed and the result is:
\begin{eqnarray}
a_{0}&=&\frac{12bz^{2}(m+1)\sqrt{6}(\mu+a_{0})}{\pi(6\mu+6a_{0}-6\kappa^{2}a_{1}-\kappa^{2}l^{2})\sqrt{\mu
      l^{2}+6\mu
      a_{1}+a_{0}l^{2}+6a_{0}a_{1}}}\arctan\left(\frac{\kappa(l^{2}+6a_{1})}{\sqrt{6(\mu+a_{0})(l^{2}+6a_{1})}}\right) \\ \nonumber
  &-&\frac{3bz^{2}(m+1)\kappa}{6\mu+6a_{0}-6\kappa^{2}a_{1}-\kappa^{2}l^{2}}\\
  \nonumber
&+&\frac{2z_{\rm c}(\mu+a_{0})^{\frac{3}{2}m+\frac{1}{2}}}{2^{3(m+1)}\pi^{\frac{3}{2}(m+1)}\Gamma\left(\frac{3}{2}(m+1)\right)\left(\frac{l^{2}}{6}+a_{1}\right)^{\frac{3}{2}(m+1)}}\left[\kappa\sqrt{\frac{l^{2}/6+a_{1}}{\mu+a_{0}}}\left(1-3m+3m\ln\left(\kappa\sqrt{\frac{l^{2}/6+a_{1}}{\mu+a_{0}}}\right)\right)\right.
\\ \nonumber
&-&\left.\arctan\left(\kappa\sqrt{\frac{l^{2}/6+a_{1}}{\mu+a_{0}}}\right)-\frac{3m\pi}{4}\ln\left(1+\frac{\kappa^{2}(l^{2}/6+a_{1})}{\mu+a_{0}}\right)\right.
\\ \nonumber
&+&\left.\frac{3mi}{2}\left(\mbox{dilog}\left(i\kappa\sqrt{\frac{l^{2}/6+a_{1}}{\mu+a_{0}}}\right)-\mbox{dilog}\left(i\kappa\sqrt{\frac{l^{2}/6+a_{1}}{\mu+a_{0}}}\right)\right)\right]\label{l6}
\end{eqnarray}
where $\mbox{dilog}(x)$ is the dilogarithmic function, defined as:
\begin{eqnarray}
\mbox{dilog}(x)=\int_{1}^{x}\mbox{d}t\,\frac{\ln(t)}{1-t}\label{l7}
\end{eqnarray}
Starting from expression 
(28) we neglect terms of order
${\cal O}(m^{2})$. As we consider weakly charged networks we only take into
account terms up to the order of $z^{2}$. Terms of higher order in the charge
$z$ are neglected. We are interested in the long ranged limit of the
Debye-H\"uckel potential. So the next step is to make a series expansion with
respect to small $\kappa$ including terms of order $\kappa^{3}$. Note, that
the sequence of the series expansions with respect to the monomer charge $z$
and the inverse Debye-H\"uckel screening length $\kappa$ is important. For
physical reasons it does not make sense to consider first the long ranged
limit before calculating the weakly charged case. The result of this analysis
is:
\begin{eqnarray}
a_{0}&=&\frac{bz^{2}(m+1)\kappa}{\mu+a_{0}}\left(\frac{2}{\pi}-\frac{1}{2}-\frac{\kappa^{2}l^{2}}{12(\mu+a_{0})}+\frac{2\kappa^{2}l^{2}}{9\pi(\mu+a_{0})}\right)\\
\nonumber
&+&\frac{\kappa^{3}z_{\rm c}}{\pi^{2}(\mu+a_{0})}\left(\frac{m\gamma}{4}+\frac{m\ln(l\kappa/\sqrt{\pi})}{2}-\frac{2m}{3}+\frac{1}{6}\right)+{\cal
  O}(m^{2},z^{4},\kappa^{4})\label{l8}
\end{eqnarray}
with $\gamma\approx 0.5772157$ Euler's constant. Neglecting terms of order
$a_{0}^{2}$ in Eq. (\ref{l8}) leads to a linear equation in $a_{0}$ which can
be solved. The result for $a_{0}$ is:
\begin{eqnarray}
a_{0}&=&\frac{\kappa}{\mu}\left(\frac{2bz^{2}(m+1)}{\pi}-\frac{bz^{2}(m+1)}{2}+\frac{2\kappa^{2}l^{2}bz^{2}(m+1)}{9\mu
    \pi}-\frac{\kappa^{2}l^{2}bz^{2}(m+1)}{12\mu}\right)\\ \nonumber
&-&\frac{\kappa}{\mu}\left(\frac{2\kappa^{2}z_{\rm
      c}m}{3\pi^{2}}-\frac{\kappa^{2}z_{\rm c}\gamma
    m}{4\pi^{2}}-\frac{\kappa^{2}z_{\rm c}\ln(\kappa
    l/\sqrt{\pi})m}{2\pi^{2}}-\frac{\kappa^{2}z_{\rm c}}{6\pi^{2}}\right)\label{l9}
\end{eqnarray}

Since the expansion coefficients $a_{0}$ and $a_{1}$ for the proper self
energy $M({\bf \hat{k}})$ due to the calculation above are known approximately
it is possible to write down the Greens function formally:
\begin{equation}
\tilde{\cal G}({\bf \hat{k}},\mu,z_{\rm c})=\frac{1}{\mu+a_{0}+\frac{l^{2}}{6}{\bf
    \hat{k}}^{2}+a_{1}{\bf \hat{k}}^{2}}\label{l11}
\end{equation}

\section{Results}
From Eq. (\ref{l11}) the grand canonical partition function in replica space
under the influence of an external constant force $f$ acting on the ends of
the chain can be calculated:
\begin{eqnarray}
\Xi_{m}(\mu,z_{\rm c},{\bf f})&=&\tilde{\cal G}({\bf k}^{(0)},{\bf k}^{(1)}-i\beta
{\bf f},\dots,{\bf k}^{(m)}-i\beta
{\bf f},\mu,z_{\rm c})\arrowvert_{{\bf \hat{k}}={\bf 0}}\nonumber \\
&=&\frac{1}{\mu+a_{0}-\frac{l^{2}}{6}m\beta^{2}{\bf f}^{2}-a_{1}m\beta^{2}{\bf
    f}^{2}}\label{erg1}
\end{eqnarray}
Here we reintroduced the force ${\bf f}$ according to Eq. (\ref{ft1a}). The monomer
chemical potential and the fugacity of cross-links which parameterize the grand
canonical partition function in Eq. (\ref{erg1}), should be expressed in terms
of the parameters $N_{\rm tot}$ and $N_{\rm c}$. According to Panyukov and Rabin
\cite{PanRab} the expression $F_{m}(N_{\rm tot},N_{\rm c})$ can be calculated by the
method of steepest descent in the thermodynamic limit $N_{\rm tot}, N_{\rm c}\to
\infty$:
\begin{eqnarray}
F_{m}(N_{\rm tot},N_{\rm c})/k_{\rm B}T=-\ln\Xi(\mu,z_{\rm c})-N_{\rm
  tot}\mu+N_{\rm c}\ln z_{\rm c}\label{erg2}
\end{eqnarray}
Consequently the fugacity $z_{\rm c}$ of cross-links and the chemical potential
$\mu$ of monomers can be obtained by minimizing the right-hand side of Eq. (\ref{erg2}): 
\begin{eqnarray}
N_{\rm tot}=-\frac{\partial\ln\Xi_{m=0}(\mu,z_{\rm c})}{\partial\mu}\label{erg4}
\end{eqnarray}
and
\begin{eqnarray}
N_{\rm c}=\frac{\partial\ln\Xi_{m=0}(\mu,z_{\rm c})}{\partial\ln z_{\rm c}}\label{erg5}
\end{eqnarray}
in the limit of a vanishing replica index $m$. It can be shown from
Eqs. (\ref{erg4}) and (\ref{erg5}), that:
\begin{eqnarray}
\bar{N}=\frac{3z_{\rm c}\mu^{2}\kappa^{3}}{8\kappa^{3}z^{2}bl^{2}\pi+36\kappa\mu
  bz^{2}\pi+3\kappa^{3}z_{\rm c}\mu-18\mu^{3}\pi^{2}-9\kappa\mu
  bz^{2}\pi^{2}-3\kappa^{3} bz^{2}l^{2}\pi^{2}}\label{erg6}
\end{eqnarray}

where $\bar{N}$ is defined as the crosslink-density $N_{\rm c}/N_{\rm tot}$ of the
network. Neglecting the charge per monomer $z$ the crosslink density reads:
\begin{eqnarray}
\bar{N}\approx\frac{3z_{\rm c}\mu^{2}\kappa^{3}}{3z_{\rm c}\mu\kappa^{3}-18\pi^{2}\mu^{3}}\label{erg7}
\end{eqnarray}
Since $\bar{N}$ is a positive number $z_{\rm c}$ has to be large enough, namely:
\begin{eqnarray}
z_{\rm c}>\frac{6\pi^{2}\mu^{2}}{\kappa^{3}}\label{erg8}
\end{eqnarray}
At this point we make a series expansion of Eq. (\ref{erg7}) with respect to
small $\mu$ neglecting terms of order $\mu^{2}$ which is possible, because
$\mu$ is connected with the crosslink-density $\bar{N}$, that is assumed to be
small:
\begin{eqnarray}
\bar{N}=\mu+{\cal O}(\mu^{2})\label{erg9}
\end{eqnarray}
To calculate $z_{\rm c}$ we make the following {\it ansatz} according to inequality
(\ref{erg8}) with a positive $y$:
\begin{eqnarray}
z_{\rm c}=\frac{6\pi^{2}\mu^{2}}{\kappa^{3}}+y\label{erg10}
\end{eqnarray}
Now we substitute this {\it ansatz} for $z_{\rm c}$ and the result for the chemical
potential $\mu$ into Eq. (\ref{erg6}) and calculate $y$. Therefore we again
make a series expansion with respect to small monomer charges $z$ neglecting
terms of order $z^{4}$. 
\begin{eqnarray}
\bar{N}&=&\frac{2\pi^{2}\bar{N}^{2}(9b\pi^{2}\bar{N}-36b\bar{N}\pi)z^{2}}{y^{2}\kappa^{5}}+\frac{2\pi^{2}\bar{N}^{2}}{y\kappa^{3}}\left(3\bar{N}+\frac{3bz^{2}l^{2}\pi^{2}-8bz^{2}l^{2}\pi}{y}\right)\\
\nonumber
&+&\frac{3bz^{2}\bar{N}\pi^{2}-12bz^{2}\bar{N}\pi}{y\kappa^{2}}+\bar{N}+\frac{3bz^{2}l^{2}\pi^{2}-8bz^{2}l^{2}\pi}{3y}+{\cal
  O}(z^{4})\label{erg11}
\end{eqnarray}
This equation can be solved with respect to $y$. Again neglecting terms of
order $z^{4}$ and only taking into account the leading term with respect to a
small inverse Debye-H\"uckel screening length $\kappa$ the result is:
\begin{eqnarray}
y=\frac{3\pi(4-\pi)bz^{2}}{\kappa^{2}}+{\cal O}(z^{4},\kappa^{0})\label{erg12}
\end{eqnarray}
Now we can write down the result for $z_{\rm c}$ which reads:
\begin{eqnarray}
z_{\rm c}=\frac{6\pi^{2}\bar{N}^{2}}{\kappa^{3}}+\frac{3\pi(4-\pi)bz^{2}}{\kappa^{2}}\label{erg13}
\end{eqnarray}
 Moreover it can be shown
\cite{PanRab}, that the conformational free energy of the network is given by:
\begin{equation}
F(N_{\rm tot},N_{\rm c},{\bf f})=-k_{\rm
  B}T\frac{\partial\ln\Xi_{m}(\mu,z_{\rm c})}{\partial m}{\Bigg\vert}_{m=0}\label{erg40}
\end{equation}
If the free energy $F$ is known the force-size relationship is the simply 
calculated by
the derivative of $F$ with respect to the external force $f$:
\begin{equation}
\langle R \rangle
=-\frac{\partial F(N_{\rm tot},N_{\rm c},f)}{\partial f}\label{erg50}
\end{equation}
After having inserted the results for $z_{\rm c}$ and $\mu$ in the force size
relationship, expanded for small charges $z$ neglecting terms of order 
$z^{4}$ and higher, since only weakly charged networks are stable, and
considered only terms of leading order with respect to small $\kappa$ the
force-size relationship reads:
\begin{equation}
\langle R \rangle=\left(\frac{bz^{2}}{3\bar{N}^{2}\pi\kappa}+\frac{l^{2}}{6\bar{N}}\right)\beta f+{\cal O}(z^{4},\kappa)\label{erg14}
\end{equation}
This is again the result for the small deformation regime. The result
describes a Hookian law for the force extension relation and defines the
elastic modulus of the polyelectrolyte network.
Note further
that this result is valid for small forces, i.e. $\beta f/\kappa<1$. 
Therefore the modulus for the small screening and the low deformation
regime of the network reads:
\begin{equation}
G=\left(\frac{\beta bz^{2}}{3\bar{N}^{2}\pi\kappa}+\frac{\beta l^{2}}{6\bar{N}}\right)^{-1}\label{erg15}
\end{equation}
which is the central result of our paper. The modulus depends on the density
of the crosslink and on the Debye screening parameter. Thus both contributions
enter in a significant way. 
Most striking is  that part of the modulus stemming from the interactions,
which  
depends on the crosslink density $\bar{N}$ squared.

\section{Discussion}

In the previous sections we analyzed the force size relationship of a
polyelectrolyte network, which was made of a very long crosslinked chain. The
method presented, which was developed for a single chain \cite{HaWiVi}, is in
replica-space applicable to polyelectrolyte networks. We considered the
network in the long ranged limit of the Debye-H\"uckel potential.

An important result is that for small forces and weakly charged networks the response to an external force ${\bf f}$ is
proportional to ${\bf f}$. The modulus $G$ depends on $\bar{N}^{2}$, where
$\bar{N}$ is the crosslink-density, which is a
surprising result in contrast to classical considerations on networks, where
the modulus is proportional to $\bar{N}$ \cite{Ltreloar}. To discuss the
result for the modulus in more detail let us discuss it in the form
\begin{equation}
G^{-1} = \beta 
\left(\frac{l^{2}}{6 \bar N}+ \frac{b z^{2}}{3 {\bar N}^{2} \pi \kappa}
\right)
\end{equation}
The modulus consists of two terms. The first part, 
$G_{\rm N} \propto k_{\rm B}T \bar N$, of the modulus 
is  the term corresponding to
classical rubber elasticity. It is proportional to the temperature and to the
crosslink density. This corresponds to the usual entropy elasticity of connected
flexible chains.
The factor $1/6$  appears only from the choice of the
special representation of the network, i.e., one macroscopic chain and has no
specific physical meaning. The second part, i.e.,
$G_{\rm I} \propto (k_{\rm B} T \bar N^{2} \kappa) /(b z^{2})$
 stems purely from the
interactions. It is not entropy elastic, since the Bjerrum length $b$ and the
Debye screening parameter depends on temperature. The overall temperature
dependence is given by $G_{\rm I} \propto T^{3/2}$. Since both parts have a
really distinguished temperature dependence, they can be separated
experimentally in a clear way. Moreover the strong difference in the crosslink
dependence, $G_{\rm N} \propto \bar N$, and $G_{\rm I} \propto \bar N^{2}$
allows also a clear experimental separation. 

It is interesting to note that the two different terms combine in as two
springs in series, 
one entropic one (the rubber network) and another energetic one, 
coming from the
interactions. If the strength of the springs is very different, 
naturally in such  systems always the weaker dominates the main
elasticity.

It is important to realize that interactions and 
elasticity interplay in a clear
way. Our results yield then the conclusion that the Flory assumption, i.e.,
adding the different parts of elastic and interaction part, is no longer
valid in these systems. These approximations are perhaps on a level of the
random phase approximation, but clearly the field theoretical
variational technique used in the
present paper is beyond perturbative methods used so far in neutral  
\cite{ball,rep} and polyelectrolyte \cite{muthy} networks. The Flory - Rehner
assumption normally uses for the total free energy of the network
$F=F_{\rm elastic} + F_{\rm int}$, i.e., the addition of the elastic and the
interaction part of the free energy \cite{flory,fh}. In this hypothesis the
{\em bare} expression of the elastic free energy is used which is proportional
to the crosslink density ${\bar N}$. Here we have shown that due to crossterms
the elastic modulus becomes renormalized by the interactions. Thus we claim
that the simple addition theorem is no longer valid.

The next step will be to apply the results from the present paper to several
experimental situations. In a subsequent and less detailed
paper \cite{preprint} we will study
the effects on swelling and the thermodynamic behavior. In following works we
will also study the free energy functional at smaller length scales which will
provide informations on the scattering behavior of the network. This is
especially important when deformation processes on different scales are
considered. Extensions to entangled systems will also become important within
this context. And finally, we have to revisit the Debye - H\"uckel
 approximation. So far,
we had assumed that the counterion are freely distributed, an assumption which
corresponds to the Debye - H\"uckel approximation \cite{borsa}. 
Of course, correlation
effects will change the picture and these higher order effects must be the
subject of subsequent studies. 

\section*{Acknowledgments}

The authors wish to thank 
 Firma Stockhausen Gmbh, D-47705 Krefeld, Germany for financial support.

\end{document}